\documentclass[aps,prl, showpacs, twocolumn, superscriptaddress,reprint]{revtex4-1}
\usepackage{amsmath,amssymb}
\usepackage{graphicx}

\begin{document}
	
	\title{Giant photoelasticity of polaritons for detection of coherent phonons in a superlattice with quantum sensitivity}
	
\author{Michal Kobecki}
	\affiliation{Experimentelle Physik 2, Technische Universit\"at Dortmund, D-44227 Dortmund, Germany}
		
\author{Alexey V. Scherbakov}
	\affiliation{Experimentelle Physik 2, Technische Universit\"at Dortmund, D-44227 Dortmund, Germany}
	\affiliation{Ioffe Institute, Russian Academy of Sciences, 194021 St. Petersburg, Russia}

\author{Serhii M. Kukhtaruk}
     \affiliation{Experimentelle Physik 2, Technische Universit\"at Dortmund, D-44227 Dortmund, Germany}
     \affiliation{Department of Theoretical Physics, V.E. Lashkaryov Institute of Semiconductor Physics, 03028 Kyiv, Ukraine}

\author{Dmytro D. Yaremkevich}
    \affiliation{Experimentelle Physik 2, Technische Universit\"at Dortmund, D-44227 Dortmund, Germany}
	
\author{Tobias Henksmeier}
    \affiliation{Department Physik, Universit\"at Paderborn, 33098 Paderborn, Germany}

\author{Alexander Trapp}
\affiliation{Department Physik, Universit\"at Paderborn, 33098 Paderborn, Germany}

\author{Dirk Reuter}
    \affiliation{Department Physik, Universit\"at Paderborn, 33098 Paderborn, Germany}

\author{Vitalyi E. Gusev}
    \affiliation{Laboratoire d'Acoustique de l'Uiversit\'e du Mans (LAUM), UMR 6613, Institut d'Acoustique - Graduate School (IA-GS), CNRS, Le Mans Universit\'e, Le Mans, France}

\author{Andrey V. Akimov}
    \affiliation{School of Physics and Astronomy, University of Nottingham, Nottingham NG7 2RD, UK}

\author{Manfred Bayer}
		\affiliation{Experimentelle Physik 2, Technische Universit\"at Dortmund, D-44227 Dortmund, Germany}
        \affiliation{Ioffe Institute, Russian Academy of Sciences, 194021 St. Petersburg, Russia}

	\begin{abstract}
The functionality of phonon-based quantum devices largely depends on the efficiency of interaction of phonons with other excitations. For phonon frequencies above 20 GHz, generation and detection of the phonon quanta can be monitored through photons. The photon-phonon interaction can be enormously strengthened by involving an intermediate resonant quasiparticle, e.g. an exciton, with which a photon forms a polariton. In this work, we discover a giant photoelasticity of exciton-polaritons in a short-period superlattice and exploit it for detecting propagating acoustic phonons. We demonstrate that 42 GHz coherent phonons can be detected with extremely high sensitivity in the time domain Brillouin oscillations by probing with photons in the spectral vicinity of the polariton resonance.
	\end{abstract}
	
		\maketitle

Recently coherent acoustic phonons with frequencies much higher than 1 GHz have been demonstrated to be prospective in quantum technologies and nanophononics ~\cite{1,2,3,4,5,6} due to their small wavelength which is comparable to the size of quantum nanodevices. Single localized phonon quanta are generated and detected using suspended nanostructures ~\cite{1,7,8,9,10,11,12} and propagating coherent phonons are suggested to become a logistic element in quantum computer networks ~\cite{13,14,15,16}. Coherent phonons with frequency higher than 20 GHz can be excited and detected exclusively using optical techniques exploiting the photon-phonon interaction which governs the conversion of phonon to photon and vice versa. Its strength is the key factor determining the efficiency and energy consumption for interconversion into coherent phonons. There are already significant achievements in the efficient generation and detection of localized coherent phonons using non-suspended optomechanical nanoresonators ~\cite{17,18,19,20,21,22}. However, for propagating phonons the sensitivity of optical methods has remained far from being able to count phonon quanta. The strength of photon-phonon coupling may be increased in non-cavity nanostructures hosting polariton resonances in which the photoelasticity increases drastically. An example is the exciton-polariton resonance in a short period semiconductor superlattice (SL) ~\cite{23,24,25}.

In the present Letter, we perform picosecond pump-probe experiments, in which we exploit an exciton-polariton resonance for detection of propagating coherent phonons with frequencies of $\sim{42}$ GHz. The coherent phonon wavepacket propagating through a short-period SL is probed by measuring the reflectivity of picosecond optical pulses with photon energy in the vicinity of the polariton resonance. The measurements show that polaritons possess giant photoelasticity and provide a three orders of magnitude higher sensitivity for detection of propagating coherent phonons than when probing apart from the polariton resonance. We show that the giant sensitivity of optical reflection to phonon associated dynamical strain owned by the polariton resonance is sufficient for detection of single phonon quanta in pump-probe setups.

The scheme of the experiment is presented in Fig.1(a). The studied SL grown on a GaAs substrate consists of 30 periods of GaAs and AlAs layers with thicknesses of 12 and 14.2 nm, respectively. The reflectivity spectrum $R_0(\hbar\omega)$ [solid line in in Fig. 1(b)] clearly shows the polariton resonance centered at $\hbar\omega_0=1.55$ eV. A wavepacket of coherent acoustic phonons is generated using pulsed optical excitation of the Al film deposited on the substrate backside. The film is excited by the pump laser pulses from a Ti-Sapphire regenerative amplifier (100-kHz repetition rate, pulse duration of 200 fs, and central photon energy of 1.55 eV). The film expands due to the optically-induced heating, and a coherent phonon wavepacket in form of a bipolar strain pulse with $\sim10$ ps duration and amplitude $\eta_0$ is injected into the GaAs substrate ~\cite{26, 27}. The typical simulated temporal profile for the used experimental scheme and materials ~\cite{28}, is shown in Fig. 1(c). The strain pulse, $\eta(t,z)$, propagates through the GaAs substrate with the velocity of longitudinal sound $\nu\approx4800$ m/s. It contains a broad spectrum of coherent longitudinal acoustic (LA) phonons,  which can be obtained by fast Fourier transform of the strain temporal profile. For the pulse shown in Fig. 1(c), it has a maximum at the frequency $f\sim20$ GHz. The experiments are performed at temperature $T=5$ K and phonons generated in the Al film reach the SL without attenuation. The coherent phonons are detected in the SL by measuring the reflectivity changes $\Delta R(t)$ of an optical probe pulse originating from the same laser.

\begin{figure}
\includegraphics[scale=1]{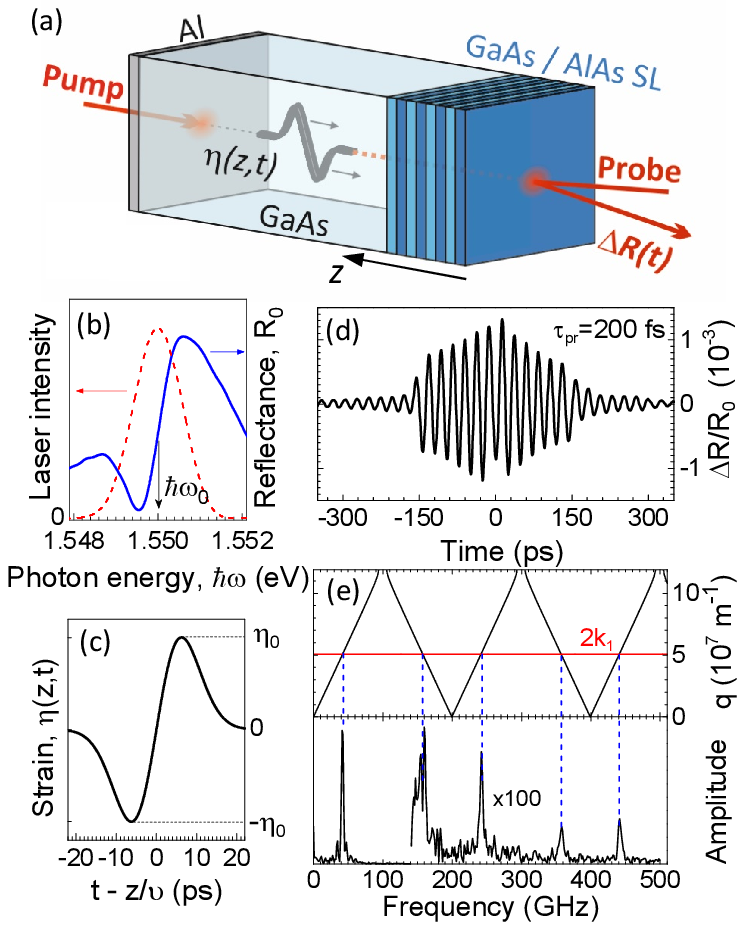}
\caption{
(a) Experimental scheme. (b) Reflectivity spectrum in the vicinity of the polariton resonance (blue curve) with subtracted background. Dashed red curve shows the spectrum of the spectrally narrow probe pulse centered at the polariton resonance.(c) Simulated spatial profile of the strain pulse injected into the GaAs substrate from the Al phonon generator.  (d) TDBS signal measured with the probe pulses of 200-fs duration. Time $t=0$ corresponds to the arrival of the phonon wavepacket center at the free surface with SL ($z$=0). (e) Folded phonon dispersion of the studied superlattice (upper panel) and the fast Fourier transform of the transient reflectivity signal shown in (d) (lower panel).
}
\end{figure}

First, we present $\Delta R(t)$ measured by the probe pulse taken straightforward from the laser system. The laser pulse with duration $\tau_{\textrm{pr}}=200$ has the spectral width of 20 meV, which covers completely the spectrum around the polariton resonance. The detected transient signal is shown in Fig. 1(d). It possesses the oscillatory behavior that is known as time domain Brillouin scattering (TDBS) ~\cite{29,30,31}. In the time interval -150 ps $\leq t \leq$ 150~ps, where the oscillations have a large amplitude, coherent phonons propagate through the SL toward the free surface, and after reflection at $t=0$ ps in the opposite direction towards the substrate. The fast Fourier transform (FFT) of $\Delta R(t)$ shown in the lower panel of Fig. 1(e) demonstrates an intense line at $f_B=42$ GHz and low intensity spectral lines with frequencies up to 450 GHz. The FFT spectrum agrees with the selection rule for TDBS $q=2k_1$ ($q$ and $k_1$ are the phonon and photon wave vectors in the SL, respectively) for normal incidence. The corresponding compliance is demonstrated in Fig.1(e), where the upper panel shows the folded dispersion relations in the studied SL ~\cite{32}. The TDBS signal in our SL is governed by phonons which frequencies are far from the SL stop bands. Therefore, phonon localization effects in the SL ~\cite{32} will not be considered further.

To study the effect of the SL polariton resonance on the TDBS signal we extend the duration of the probe pulse up to $\tau_{\textrm{pr}}=1.35$ ps with a corresponding narrowing of its spectral width down to 1.4 meV by using a tunable filter and measure the TDBS signal $\Delta R(t)$ for different central photon energy $\hbar\omega$. The spectrum of the extended probe pulse for $\hbar\omega=\hbar\omega_0$  is shown in Fig. 1(b) by the dashed red line. The value of $\hbar\omega$ is varied in the vicinity of the polariton resonance between 1.544 and 1.556 eV. In order to avoid nonlinear exciton effects ~\cite{33,34}, we keep the probe fluence on the surface with the SL to less than 300 nJ/cm2.

Figure 2(a) shows the detected signal for a number of detuning values $\hbar\omega - \hbar\omega_0$. It is seen that the amplitude of the oscillations strongly depends on the probe pulse photon energy. Within the time interval when coherent phonons propagate in the SL, the measured signal can be fit with high precision by rising ($t<0$) and decaying ($t>0$) harmonic oscillations of single frequency, $f_{\textrm{B}}$, amplitude, $A_{\textrm{B}}$, phase, $p_{\textrm{B}}$, and rise or decay rates, $\tau_{\textrm{B}}^{-1}$, respectively. The dependences of these parameters are presented by the symbols in Figs. 2(b) and 2(c): open symbols for phonons propagating toward the free surface (anti-Stokes) and filled symbols for phonons propagating in the opposite direction after reflection (Stokes). The dependences are symmetric for $A_{\textrm{B}}$ and $\tau_{\textrm{B}}^{-1}$ and antisymmetric for $f_{\textrm{B}}$ and $p_{\textrm{B}}$ relatively to $\hbar\omega_0$. This leads us to the conclusion that the TDBS signals and dependences $A_{\textrm{B}}$, $\tau_{\textrm{B}}^{-1}$,  $f_{\textrm{B}}$ and $p_{\textrm{B}}$ are governed by the polariton resonance when probing with $\hbar\omega$ close to $\hbar\omega_0$.

\begin{figure}
\includegraphics[scale=0.95]{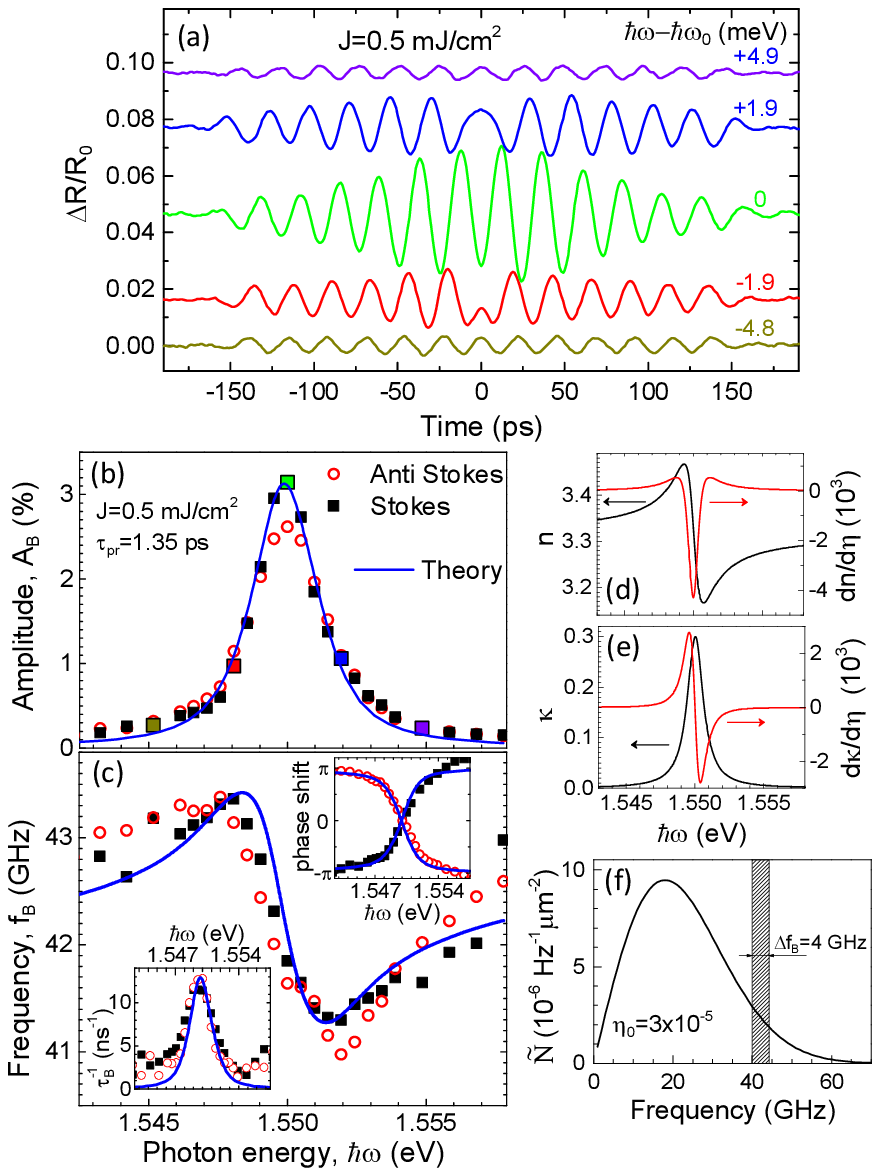}
\caption{
(a) TDBS signals measured by the spectrally narrowed probe pulses for several $\hbar\omega$. (b,c) Dependences of the specific properties of the TDBS signals on the probe photon energy: amplitude $A_{\textrm{B}}$ (b), frequency $f_{\textrm{B}}$ (c), decay rate $\tau_{\textrm{B}}^{-1}$ [lower insert in (c)], and a shift of the phase $p_{\textrm{B}}$ relatively to the signal measured at the resonant conditions [upper insert in (c)]. Enlarged colored symbols in (b) correspond to the transient signals shown by the same color in (a). (d,e) Calculated spectral dependences of the real and imaginary part of refractive index and their derivatives in the vicinity of the polariton resonance. (f) Spectral-spatial density of phonons in the coherent phonon wavepacket for $J=0.5$ mJ/cm$^2$. Shaded area shows the spectral range of detected phonons.
}
\end{figure}

The exciting experimental result is the observation of the huge amplitude of the TDBS signal. When probing coherent phonons at the polariton resonance~($\hbar\omega=\hbar\omega_0$), the relative changes $\Delta R/R_0$ are $\sim10^{-2}$ for the used pump fluence $J\sim0.1$ mJ/cm$^2$. The measurements of TDBS in a material without a narrow optical resonances for a similar wavepacket of coherent phonons would give $\Delta R/R_0 \sim 10^{-5}$ ~\cite{32,35} which is three orders of magnitude smaller than measured for the detection at the polariton resonance in the present work. This result means that our experiments reveal a giant photoelasticity of polaritons and extremely high sensitivity to propagating coherent phonons.

For qualitative analysis of the experimental results, we use a simplified model in which we assume that the spectral width of the probe pulse is much smaller than the width of the polariton resonance. In this case the amplitude of the TDBS signal can be estimated from the following equation ~\cite{36}:
    \begin{equation}
   \frac{\Delta R(t)}{R_0}=2\textrm{Re}\left[{i\frac{1-r_{01}^2}{r_{01}}\frac{dk_1}{d\eta}\int\limits_{0}^{\infty}\eta(z,t)e^{2ik_1z}dz}\right]
   \end{equation}

where $k_1=2\pi \tilde{n}/\lambda$ and $\tilde{n}$ and $\lambda$ are the complex refractive index in the SL and the wavelength of the probe light in vacuum, respectively, $r_{01}=(1-\tilde{n})/(1+\tilde{n})$ and $\eta(z,t)=\eta(z\pm \nu t)$ is the time-spatial profile of the strain pulse propagating in the SL along $z$ ($z=0$ at the free surface of SL) toward the free surface (+) and backwards (-). The crucial parameter in Eq. (1) which governs the sensitivity of detection is the derivative of $k_1$ on strain $\eta$ which is defined by the strain dependence of the effective dielectric function in the SL with the polariton resonance given by ~\cite{37}:
    \begin{equation}
   \varepsilon_{\textrm{eff}}(\omega)=\varepsilon_{\textrm{b}}\left[{1+\frac{\omega_{\textrm{LT}}}{\omega_0-\omega-i\Gamma}}\right]
    \end{equation}
where $\varepsilon_{\textrm{b}}$ is the background dielectric constant, $\omega_{\textrm{LT}}$ and $\Gamma$ are the longitudinal-transverse splitting defined by the interaction of the excitons with light and the nonradiative decay rate of the polaritons, respectively. The deformation potential mechanism is the main one responsible for the phonon-induced changes of $\varepsilon_{\textrm{eff}}(\omega)$: the strain associated with coherent phonons induces the energy shift of the polariton resonance, i.e. $\hbar\omega_0$ ~\cite{38}. If this shift is much smaller than $\Gamma$ we get
 \begin{equation}
\frac{dk_1}{d\eta}=\frac{1}{2}\frac{k_1}{\varepsilon_{\textrm{eff}}}\frac{\Xi }{\hbar}\frac{d\varepsilon_{\textrm{eff}}}{d\omega_0}.
\end{equation}
where $\Xi=-10$ eV is the deformation potential for excitons in GaAs ~\cite{39}.

The presented model explains the measured TDBS signals and dependences presented in Fig. 2(b) and 2(c). The explanation comes from the dependences of the real $(n)$ and imaginary $(\kappa)$ parts of the refractive index $\tilde{n}=\sqrt{\varepsilon_{\textrm{eff}}}$ and their derivatives on strain. They are shown in Figs. 2(d) and 2(e) for the polariton parameters, which fit our experimental data: $\hbar\omega_0=1.550$~eV, $\omega_{\textrm{LT}}=0.13$~meV and $\Gamma=0.7$~meV. It is seen that at $\hbar\omega=\hbar\omega_0=1.55$~eV, $\left|{\frac{dn}{d\eta}}\right|\approx 4\times10^3$ possesses an extremum which results in the maximum for $A_{\textrm{B}}(\hbar\omega)$ [Fig. 2(b)]. Tuning  $\hbar\omega$ away from the polariton resonance leads to the decrease of $\left|{\frac{dn}{d\eta}}\right|$  which obviously leads to the decrease of the TDBS signal amplitude observed experimentally. The dependence of $f_{\textrm{B}}$ in Fig. 2(c) qualitatively follows $n(\hbar\omega)$, in full agreement with the wavevector selection rule $q=2k_1$  for the phonon-polariton interaction which governs the TDBS signal. The measured dependence of the rise/decay rate $\tau_{\textrm{B}}^{-1}$ of the TDBS oscillations shown in the lower inset in Fig. 2(c) is similar to the dependence $\kappa(\hbar\omega)$, and the phase of the oscillations $p_{\textrm{B}}$ changes by $\pi$ in the vicinity of the polariton resonance [the upper inset in Fig. 2(c)], following the sign of $d\kappa/d\omega$.

For a quantitative analysis, we developed a comprehensive theoretical approach ~\cite{40}, which takes into account the spectral width of the probe pulse and the Stokes/anti-Stokes energy shift of the reflected light. The solid curves in Figs. 2(b) and 2(c) are the results of calculations for the parameters given above. Excellent agreement between the measured and calculated dependences is seen.

Now we turn to the discussion of phonon quantum sensitivity of polaritons. The number of phonons, $N_{\textrm{B}}$, responsible for the TDBS is determined by their spectral density $\tilde{N}$ in the phonon wavepacket around the Brillouin frequency $f_{\textrm{B}}$. It is determined by the spatial-temporal shape of the generated phonon wavepacket $\eta(z,t)$. Fig. 2(f) is an example of the phonon spectral density ~\cite{41} for the strain pulse shown in Fig. 1(c) with the amplitude $\eta_0=3\times10^{-5}$, which corresponds to $J=0.5$ mJ/cm$^2$ ~\cite{35}. The shaded area centered at the frequency $f_{\textrm{B}}=42$ GHz indicates the spectral range of phonons detected at  $\hbar\omega=\hbar\omega_0$. The finite spectral width, $\Delta f_{\textrm{B}}$, is due to the finite size of the SL and the penetration depth of light, which determines the rise/decay rate of the TDBS signal. For $\tau^{-1}_{\textrm{B}}=12$ ns$^{-1}$, $\Delta f_{\textrm{B}}=4$ GHz and the estimated density of detected phonons $N_{\textrm{B}}\approx10^{4}$ $\mu$m$^{-2}$. Let us assume the prospective case when we need to detect phonon quanta emitted by a semiconductor nanodevice, e.g. a quantum dot ~\cite{42,43}. A phonon wavepacket containing one phonon quantum of the Brillouin frequency per ~1-$\mu$m$^2$ detected in the probe spot of the corresponding diameter will induce a TDBS signal with amplitude $\sim10^{-6}$. Such a signal can be reliably detected with high repetition rate pump-probe setups.

\begin{figure}
\includegraphics[scale=1]{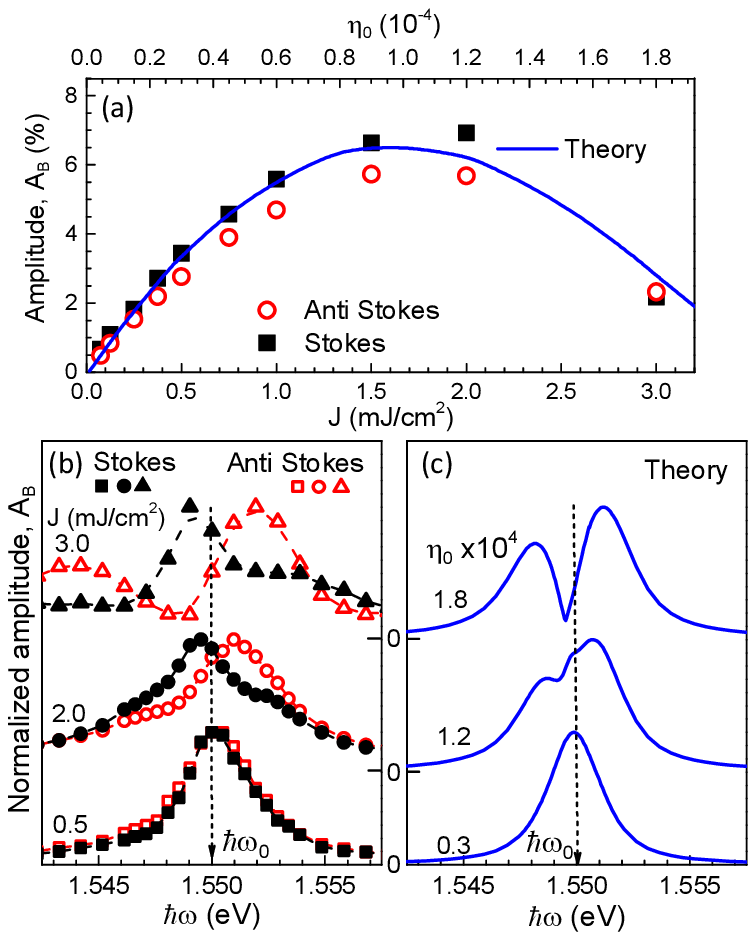}
\caption
{
(a) Dependence of the TDBS signal amplitude on the optical excitation density (experimental data, symbols) and the strain pulse amplitude (theoretical calculations, solid line). The experiential errors do not exceed the symbols size. (b,c) Measured (b) and calculated (c) dependences of the TDBS signal amplitude on the probe photon energy for three values of $J$ and $\eta_0$, respectively. The dependences are normalized to the values at the maxima.
}
\end{figure}

The experimental results and theoretical consideration presented above concern a small number of coherent phonons where the phonon-induced shift of the exciton resonances in the SL quantum wells is negligibly small. In this regime, the amplitude $A_{\textrm{B}}$ of the TDBS oscillations depends linearly on the number of phonons with $f=f_{\textrm{B}}$ and, thus, linearly on $J$. This consideration is likely not valid for high pump fluence when the maximum strain $\eta_0$ in the coherent phonon wavepacket induces the exciton shift $\Delta\hbar\omega=\eta_0\Xi$ by a value comparable to the width $\Gamma$ of the polariton resonance. Then $\frac{dk_1}{d\eta}$ given by Eq. (3) becomes dependent on time and coordinate. Qualitatively, at high $J$ the polariton resonance broadens and at $\hbar\omega=\hbar\omega_0$ the sensitivity to phonons of the Brillouin frequency decreases. Indeed, the dependence $A_{\textrm{B}}(J)$ measured at $\hbar\omega=\hbar\omega_0$ and shown in Fig. 3(a) by the symbols saturates at $J\approx2$ mJ/cm$^2$, and a further increase of $J$ results in a decrease of $A_{\textrm{B}}$. The measured dependence perfectly agrees with the theoretical simulations [solid line in Fig. 3(a)] for the strain pulse shape shown in Fig. 1(c). We have checked that the observed nonlinearity is not related to nonlinear acoustic effects ~\cite{44,45} which start to become pronounced in our experiment at $J>3$~mJ/cm$^2$.

Nonlinear photoelastic effects emerge also in the experimental and theoretical dependences $A_{\textrm{B}}(\hbar\omega)$ for high $J$ as shown in Figs. 3(b) and 3(c). It is seen that at moderate $J = 2$ mJ/cm$^2$, and high $J=3$ mJ/cm$^2$ fluence the experimentally measured spectral shape of $A_{\textrm{B}}(\hbar\omega)$ [Fig. 3(b)] shifts by ~1 meV relative to $\hbar\omega_0$ and this shift has opposite signs for the Stokes and Anti Stokes signals. The maxima in the theoretical curves [Fig. 3(c)] also shifts, but the curves are the same for the Stokes and Anti Stokes cases contrary to the experimental results. This difference between experiment and theory can be explained by the asymmetric temporal shape of the strain pulse with a predominant compression component ~\cite{28}. In this case, the phonons propagating towards the free surface of the SL (Anti Stokes) induce the blue shift of the exciton resonance. After reflection at the free surface the tensile component of the strain pulse becomes dominant inducing the red shift of the exciton resonance. In the theoretical simulations such asymmetry is not included and the calculated dependence $A_{\textrm{B}}(\hbar\omega)$ demonstrates a spectral shape independent on the phonon propagation direction.

In conclusion, we have demonstrated the effect of giant polariton photoelasticity for detecting propagating coherent phonons by time domain Brillouin scattering. Our results pave a path for prospective exploitation of resonant polariton excitations for manipulating phonons on the quantum level. The strong dispersion of the permittivity in the visible range in the vicinity of the polariton resonance results in a huge ultrafast response of the optical properties to dynamical strain which accompanies the coherent phonons. We have developed a quantitative theoretical model which allows us to predict the absolute values for the optical reflectivity change induced by the propagating phonons with Brillouin frequency. We have demonstrated that nonlinear effects for a boosted density of phonon flux suppress the sensitivity, in full agreement with the deformation potential model for exciton-phonon interactions used in treating the observed phenomena. The frequency of detected phonons in our experiment is determined by the velocity of sound and the refractive index of the SL at the probe photon energy, which is set by the spectral position of the exciton  resonance. It can be adjusted by changing the widths of the GaAs/AlAs layers and by using alternative combinations of materials, such as nitrides ~\cite{46} or II-VI semiconductors ~\cite{47}. Moreover, the folded phonon dispersion provides access to the detection of phonons with THz frequencies \cite{24}. This opens great perspectives for optical detection of single high-frequency phonon quanta via exciton polaritons.

We are grateful to Ilya Akimov, Nikolay Gippius, Mark A${\ss}$mann and Christine Silberhorn for fruitful discussions. The work was supported by the Deutsche Forschungsgemeinschaft (Grant No. TRR 142 project A06 and Grant No. TRR160 project A01).


\begin{thebibliography}{200}
\bibitem {41} The spectral density of phonons was calculated by normalizing the phonon spectral power density obtained by fast Fourier transform of $\eta(z,t)$ by the total energy of the strain pulse: $U=\frac{1}{2} V c_{11} \langle\eta^2\rangle\approx10^{-14}$ J, where $V=\pi r^2 l$ is the strain pulse total volume ($r=27$ $\mu$m is the pump spot diamter and $l=200$ nm is the strain pulse spatial size along the $z$-axis), $c_{11}=119$ GPa is the GaAs elastic constant and the averaged strain amplitude $\langle\eta^2\rangle$ is $3.25\times10^{-10}$ for $J=0.5$ mJ/cm$^2$ ~\cite{35}.
\end{thebibliography}

\end{document}